\documentclass[10pt]{article}

\newlength{\minitwocolumn}
\font\teneufm=eufm10
\font\seveneufm=eufm7
\font\fiveeufm=eufm5
\newfam\eufmfam
\textfont\eufmfam=\teneufm
\scriptfont\eufmfam=\seveneufm
\scriptscriptfont\eufmfam=\fiveeufm

\makeatletter
\@addtoreset{equation}{section}
\makeatother

\title{\bf
\Large{\bf 
BOSONIZATION OF
SUPERALGEBRA $U_q(\widehat{sl}(N|1))$\\
FOR AN ARBITRARY LEVEL
}}
\begin{document}

\maketitle

\begin{center}
{Takeo Kojima}

{\it
Department of Mathematics and Physics,
Faculty of Engineering,
Yamagata University,\\
 Jonan 4-3-16, Yonezawa 992-8510, JAPAN\\
kojima@yz.yamagata-u.ac.jp}
\end{center}

~\\

\begin{abstract}
We give a bosonization of the quantum affine superalgebra
$U_q(\widehat{sl}(N|1))$ for an arbitrary level $k \in {\bf C}$.
The bosonization of level $k \in {\bf C}$ is completely different from
those of level $k=1$.
From this bosonization, 
we induce the Wakimoto realization whose character coincides with
those of the Verma module.
We give the screening
that commute with $U_q(\widehat{sl}(N|1))$.
Using this screening,
we propose the vertex operator that is the intertwiner among 
the Wakimoto realization and typical realization.
We study non-vanishing property of the correlation function
defined by a trace of the vertex operators. 
\end{abstract}

\newpage

\section{Introduction}
Bosonizations provide a powerful method to 
construct correlation function of exactly solvable models.
We construct a bosonization of the quantum affine superalgebra
$U_q(\widehat{sl}(N|1))$ $(N \geq 2)$ 
for an arbitrary level $k \in {\bf C}$ \cite{AOS2, K1}.
For the special level $k=1$,
bosonizations have been constructed for the quantum affine algebra
$U_q(g)$ in many cases $g=(ADE)^{(r)}$,
$(BC)^{(1)}$, $G_2^{(1)}$, $\widehat{sl}(M|N)$, $osp(2|2)^{(2)}$
\cite{FJ, B, JKM, J1, J2, KSU, Z, YZ}.
Bosonizations of level $k \in {\bf C}$ are completely different from
those of level $k=1$.
For an arbitrary level $k \in {\bf C}$
bosonizations have been studied only for $U_q(\widehat{sl}_N)$ 
\cite{S, AOS1} and $U_q(\widehat{sl}(N|1))$ \cite{AOS2, K1}.
Our construction is based on the ghost-boson system.
We need more consideration to get the Wakimoto realization
whose character coincides with
those of the Verma module.
Using $\xi$-$\eta$ system 
we construct the Wakimoto realization \cite{ZG, K2}
from our level $k$ bosonization.
For an arbitrary level $k \neq -N+1$
we construct the screening current that commutes with 
$U_q(\widehat{sl}(N|1))$ modulo total difference.
By using Jackson integral and the screening current,
we construct the screening that commute with $U_q(\widehat{sl}(N|1))$
\cite{ZG, K3}.
We propose the vertex operator that is the intertwiner among 
the Wakimoto realization and typical realization.
By using the Gelfand-Zetlin basis,
we have checked the intertwining property of the vertex operator
for rank $N=2,3,4$ \cite{K3}.
We balance the background charge of the vertex operator
by using the screening and propose the correlation function
by a trace of them, which gives quantum and super generalization of 
Dotsenko-Fateev theory \cite{DF}.

The paper is organized as follows.
In section 2 we review bosonizations of $U_q(\widehat{sl}_2)$.
In section 3 we construct a bosonization 
of $U_q(\widehat{sl}(N|1))$ for an arbitrary level $k \in {\bf C}$.
We induce the Wakimoto realization by $\xi$-$\eta$ system.
In section 4 we construct the screening that commute 
with $U_q(\widehat{sl}(N|1))$ for an arbitrary
level $k \neq -N+1$.
We propose the vertex operator and the correlation function.

\section{Bosonization : Level $k=1$ vs. Level $k \in {\bf C}$}

In this section we review the bosonization of
the quantum affine algebra $U_q(\widehat{sl}_2)$.
The purpose of this section is to make
readers understand that the bosonization of level 
$k \in {\bf C}$ is complete different from those of level $k=1$.
In what follows let $q$ be a generic complex number $0<|q|<1$.
We use the standard $q$-integer notation :
\begin{eqnarray}
[m]_q=\frac{q^m-q^{-m}}{q-q^{-1}}.\nonumber
\end{eqnarray}
First we recall the definition of $U_q(\widehat{sl}_2)$.
We recall the Drinfeld realization of the quantum affine algebra
$U_q(\widehat{sl}_2)$.

~\\
{\bf Definition 2.1}
{\it~\cite{D}~
The generators of the quantum affine algebra $U_q(\widehat{sl}_2)$
are
$x_{i,n}^\pm$, $h_m$, $h$, $c$~$(n \in {\bf Z}, m \in {\bf Z}_{\neq 0})$.
Defining relations are
\begin{eqnarray}
&&c : {\rm central},~[h, h_m]=0,\nonumber
\\
&&[h_m,h_n]=\delta_{m+n,0}\frac{[2 m]_q[c m]_q}{m},
\nonumber
\\
&&[h,x^\pm(z)]=\pm 2 x^\pm(z),
\nonumber\\
&&[h_m,x^\pm(z)]=\pm \frac{[2 m]_q}{m}q^{\mp \frac{c|m|}{2}}z^m x^\pm(z),
\nonumber\\
&&(z_1-q^{\pm 2}z_2)x^\pm(z_1)x^\pm(z_2)=
(q^{\pm 2}z_1-z_2)x^\pm(z_2)x^\pm(z_1),
\nonumber\\
&&[x^+(z_1),x^-(z_2)]=\frac{1}{(q-q^{-1})z_1 z_2}\nonumber\\
&\times&
\left(\delta(q^{-c}z_1/z_2)\psi^+(q^{\frac{c}{2}}z_2)-
\delta(q^c z_1/z_2)\psi^-(q^{-\frac{c}{2}}z_2)
\right).\nonumber
\end{eqnarray}
where we have used $\delta(z)=\sum_{n \in {\bf Z}}z^n$.
We have set the generating function
\begin{eqnarray}
x^\pm(z)&=&\sum_{n \in {\bf Z}}x_n^\pm z^{-n-1},
\nonumber\\
\psi^\pm(q^{\pm \frac{c}{2}}z)&=&
q^{\pm h}e^{\pm (q-q^{-1})
\sum_{m>0}h_{\pm m}z^{\mp m}}.
\nonumber
\end{eqnarray}
}
When the center $c$ takes the complex number $c=k \in {\bf C}$, we call it
the level $k$ representation.
We call the realization by the differential operators the bosonization.
Frenkel-Jing \cite{FJ} constructed the level $k=1$
bosonization of the quantum affine algebra $U_q(g)$ 
for simply-laced $g=(ADE)^{(1)}$.
Here we recall the level $k=1$ bosonization of $U_q(\widehat{sl}_2)$.
We introduce the boson 
$a_n$ $(n \in {\bf Z}_{\neq 0})$
and the zero-mode operator $\partial, \alpha$ by
\begin{eqnarray}
&&[a_m,a_n]=\frac{[2m]_q [m]_q}{m}\delta_{m+n,0},~~~[\partial, \alpha]=2.
\nonumber
\end{eqnarray}
In what follows,
in order to avoid divergences, we restrict ourselves to the Fock space of the bosons. 

~\\
{\bf Theorem 2.2}
{\it~\cite{FJ}~
A bosonization of the quantum affine algebra $U_q(\widehat{sl}_2)$
for the level $k=1$ is given as follows.
\begin{eqnarray}
&&c=1,~~h=\partial,~~h_n=a_n,
\nonumber\\
&&x^\pm(z)=:e^{\mp \sum_{n \neq 0}\frac{a_n}{[n]_q}q^{\mp \frac{n}{2}}z^{-n}
\pm (\alpha+\partial)}:.
\nonumber
\end{eqnarray}
}
We have used the normal ordering symbol $: :$
\begin{eqnarray}
:a_k a_l:
=\left\{\begin{array}{cc}
a_k a_l& (k<0),\\
a_l a_k& (k>0),
\end{array}
\right.~~~:\alpha \partial:=:\partial \alpha:=\alpha \partial.
\nonumber
\end{eqnarray}

Next we recall the level $k$ bosonization of 
the quantum affine algebra $U_q(\widehat{sl}_2)$ \cite{S}.
We introduce the bosons and the zero-mode operator
$a_n, b_n, c_n$, $Q_a, Q_b, Q_c$ $(n \in {\bf Z})$ as follows.
\begin{eqnarray}
&&[a_m,a_n]=\delta_{m+n,0}\frac{[2m]_q [(k+2)m]_q}{m},~[\tilde{a}_0,Q_a]=2(k+2),
\nonumber\\
&&[b_m,b_n]=-\delta_{m+n,0}\frac{[2m]_q[2m]_q}{m},~[\tilde{b}_0,Q_b]=-4,
\nonumber\\
&&[c_m,c_n]=\delta_{m+n}\frac{[2m]_q[2m]_q}{m},~[\tilde{c}_0,Q_c]=4,
\nonumber
\end{eqnarray}
where $\tilde{a}_0=\frac{q-q^{-1}}{2 {\rm log}q}a_0$, 
$\tilde{b}_0=\frac{q-q^{-1}}{2 {\rm log}q}b_0$,
$\tilde{c}_0=\frac{q-q^{-1}}{2 {\rm log}q}c_0$.
It is convenient to introduce the generating function $a(N|z;\alpha)$.
\begin{eqnarray}
a( N |z;\alpha )=
-\sum_{n \neq 0}\frac{a_n}{[N n]_q}q^{|n|\alpha}z^{-n}
+\frac{\tilde{a}_0}{N}{\rm log}z+
\frac{Q_a}{N}.\nonumber
\end{eqnarray}
In what follows, in order to avoid divergences,
we restrict ourselves to the Fock space of the bosons.

~\\
{\bf Theorem 2.3}
{\it~\cite{S}~
A bosonization of the quantum affine algebra 
$U_q(\widehat{sl}_2)$ for the level
$k \in {\bf C}$ is given as follows.
\begin{eqnarray}
c
&=&k \in {\bf C},~~h=a_0+b_0,
\nonumber\\
h_m
&=&q^{2m-|m|}a_m+q^{(k+2)m-\frac{k+2}{2}|m|}b_m,
\nonumber\\
x^+(z)
&=&\frac{-1}{(q-q^{-1})z}\left(
:e^{-b(2|q^{-k-2}z;1)-c(2|q^{-k-1}z;0)}:\right.\nonumber\\
&&~~~~~~~~~~~~~~-
\left.:e^{-b(2|q^{-k-2}z;1)-c(2|q^{-k-3}z;0)}:\right),\nonumber
\\
x^-(z)&=&
\frac{1}{(q-q^{-1})z}\left(
:e^{a(k+2|q^k z,-\frac{k+2}{2})-a(k+2|q^{-2}z;\frac{k+2}{2})+b(2|z;-1)+c(2|q^{-1}z;0)}:\right.\nonumber\\
&&-\left.
:e^{a(k+2|q^{-k-4}z;-\frac{k+2}{2})-a(k+2|q^{-2}z;\frac{k+2}{2})
+b(2|q^{-2k-4}z;-1)+c(2|q^{-2k-3}z;0)}:
\right).\nonumber
\end{eqnarray}
}
The level $k=1$ bosonization is given by "monomial".
The level $k \in {\bf C}$ bosonization is given by "sum".
They are completely different.

\section{Bosonization of Quantum Superalgebra $U_q(\widehat{sl}(N|1))$}

In this section we study the bosonization of the quantum superalgebra
$U_q(\widehat{sl}(N|1))$ for an arbitrary level $k \in {\bf C}$.

\subsection{Quantum Superalgebra $U_q(\widehat{sl}(N|1))$}

In this section we recall the definition of
the quantum superalgebra
$U_q(\widehat{sl}(N|1))$.
We fix a generic complex number $q$ such that $0<|q|<1$.
The Cartan matrix 
$(A_{i,j})_{0\leq i,j \leq N}$
of the affine Lie algebra $\widehat{sl}(N|1)$ is
given by
\begin{eqnarray}
A_{i,j}=
(\nu_i+\nu_{i+1})\delta_{i,j}-
\nu_i \delta_{i,j+1}-\nu_{i+1}\delta_{i+1,j}.
\nonumber
\end{eqnarray}
Here we set $\nu_1=\cdots =\nu_N=+, \nu_{N+1}=\nu_0=-$.
We introduce the orthonormal basis $\{\epsilon_i|
i=1,2,\cdots,N+1\}$
with the bilinear form,
$(\epsilon_i|\epsilon_j)=\nu_i \delta_{i,j}$.
Define 
$\bar{\epsilon}_i=\epsilon_i-\frac{\nu_i}{N-1}
\sum_{j=1}^{N+1}\epsilon_j$.
Note that $\sum_{j=1}^N \bar{\epsilon}_j=0$.
The classical simple roots $\bar{\alpha}_i$
and the classical fundamental weights 
$\bar{\Lambda}_i$
are defined by
$\bar{\alpha}_i=\nu_i \epsilon_i-\nu_{i+1} \epsilon_{i+1}$,
$\bar{\Lambda}_i=\sum_{j=1}^i \bar{\epsilon}_j$
$(1\leq i \leq N)$.
Introduce
the affine weight $\Lambda_0$
and the null root
$\delta$ satisfying
$(\Lambda_0|\Lambda_0)=(\delta|\delta)=0$,
$(\Lambda_0|\delta)=1$, $(\Lambda_0|\epsilon_i)=0$,
$(\delta|\epsilon_i)=0$, $(1\leq i \leq N)$.
The other affine weights
and the affine roots
are given by
$\alpha_0=\delta-\sum_{j=1}^N \bar{\alpha}_j$,
$\alpha_i=\bar{\alpha}_i$,
$\Lambda_i=\bar{\Lambda}_i+\Lambda_0$,
$(1\leq i \leq N)$.
Let
$P=\oplus_{j=1}^N{\bf Z}\Lambda_j
\oplus {\bf Z}\delta$
and $P^*=\oplus_{j=1}^N{\bf Z}h_j
\oplus {\bf Z}d$
the affine $\widehat{sl}(N|1)$ weight lattice 
and its dual lattice, respectively.

~\\
{\bf Definition 3.1}
{\it ~\cite{Y}~
The quantum affine
superalgebra $U_q(\widehat{sl}(N|1))$ are
generated by
the generators
$h_i, e_i, f_i~(0\leq i \leq N)$.
The ${\bf Z}_2$-grading
of the generators are $|e_0|=|f_0|=|e_N|=|f_N|=1$
and zero otherwise.
The defining relations are
given as follows.
\\
{\rm The Cartan-Kac relations : }  For $N \geq 2$, $0\leq i,j \leq N$,
the generators subject to the following relations.
\begin{eqnarray}
&&[h_i,h_j]=0,~[h_i,e_j]=A_{i,j}e_j,
~[h_i,f_j]=-A_{i,j}f_j,~
[e_i,f_j]=\delta_{i,j}\frac{q^{h_i}-q^{-h_i}}{q-q^{-1}}.\nonumber
\end{eqnarray}
{\rm The Serre relations : }  For $N \geq 2$, 
the generators subject to the following relations
for $1\leq i \leq N-1$, $0\leq j \leq N$ such that $|A_{i,j}|=1$.
\begin{eqnarray}
~[e_i,[e_i,e_j]_{q^{-1}}]_q=0,~
~[f_i,[f_i,f_j]_{q^{-1}}]_q=0.\nonumber
\end{eqnarray}
For $N \geq 2$, 
the generators subject to the following relations for 
$0\leq i,j \leq N$ such that $|A_{i,j}|=0$.
\begin{eqnarray}
~[e_i,e_j]=0,~~~[f_i,f_j]=0.\nonumber
\end{eqnarray}
For $N \geq 3$, the Serre relations of fourth degree hold.
\begin{eqnarray}
\left.\begin{array}{cc}
~[e_N,[e_0,[e_N,e_{N-1}]_{q^{-1}}]_q]=0,&
~[e_0,[e_1,[e_0,e_N]_q]_{q^{-1}}]=0,\nonumber
\\
~[f_N,[f_0,[f_N,f_{N-1}]_{q^{-1}}]_q]=0,&
~[f_0,[f_1,[f_0,f_N]_q]_{q^{-1}}]=0.
\end{array}\right.\nonumber
\end{eqnarray}
For $N=2$, the extra Serre relations of fifth degree hold.
\begin{eqnarray}
\left.\begin{array}{cc}
&[e_2,[e_0,[e_2,[e_0,e_1]_q]]]_{q^{-1}}=
[e_0,[e_2,[e_0,[e_2,e_1]_{q}]]]_{q^{-1}},
\nonumber
\\
&[f_2,[f_0,[f_2,[f_0,f_1]_q]]]_{q^{-1}}=
[f_0,[f_2,[f_0,[f_2,f_1]_{q}]]]_{q^{-1}}.
\end{array}\right.\nonumber
\end{eqnarray}
Here and throughout this paper,
we use the notations
\begin{eqnarray}
~[X,Y]_\xi=XY-(-1)^{|X||Y|}\xi YX.\nonumber
\end{eqnarray}
We write $[X,Y]_1$ as $[X,Y]$ for simplicity.}\\
The quantum affine superalgebra $U_q(\widehat{sl}(N|1))$
has the ${\bf Z}_2$-graded Hopf-algebra structure.
We take the following coproduct
\begin{eqnarray}
\Delta(e_i)=e_i\otimes 1+q^{h_i}\otimes e_i,~
\Delta(f_i)=f_i\otimes q^{-h_i}+1 \otimes f_i,~
\Delta(h_i)=h_i \otimes 1+1 \otimes h_i, \nonumber
\end{eqnarray}
and the antipode
\begin{eqnarray}
S(e_i)=-q^{-h_i}e_i,~
S(f_i)=-f_i q^{h_i},~
S(h_i)=-h_i.\nonumber
\end{eqnarray}
The coproduct $\Delta$ satisfies an algebra automorphism
$\Delta(XY)=\Delta(X)\Delta(Y)$
and the antipode $S$ satisfies
a ${\bf Z}_2$-graded algebra anti-automorphism
$S(XY)=(-1)^{|X||Y|}S(Y)S(X)$.
The multiplication rule 
for the tensor product is ${\bf Z}_2$-graded 
and is defined for homogeneous elements
$X,Y,X',Y' \in U_q(\widehat{sl}(N|1))$ and
$v \in V, w \in W$ by
$X \otimes Y \cdot X' \otimes Y'=(-1)^{|Y||X'|}
X X' \otimes Y Y'$ and
$X \otimes Y \cdot v \otimes w=(-1)^{|Y||v|}
X v \otimes Y w$,
which extends to inhomogeneous elements through linearity.

~\\
{\bf Definition 3.2}{\it~~The quantum superalgebra
$U_q(\widehat{sl}(N|1))$ is the subalgebra of $U_q(\widehat{sl}(N|1))$,
that is generated by
$e_1,e_2,\cdots,e_N$, $f_1,f_2,\cdots,f_N$, and
$h_1,h_2,\cdots, h_N$.}

~\\
We recall the Drinfeld realization of $U_q(\widehat{sl}(N|1))$,
that is convenient to construct bosonizations.

~\\
{\bf Definition 3.3}{\it~\cite{Y}~
The generators of
the quantum
superalgebra $U_q(\widehat{sl}(N|1))$
are
$x_{i,n}^\pm$, $h_{i,m}$, $h$, 
$c$ $(1\leq i \leq N,
n \in {\bf Z},
m \in {\bf Z}_{\neq 0})$.
Defining relations are
\begin{eqnarray}
&&~c : {\rm central},~[h_i,h_{j,m}]=0,\nonumber
\\
&&~[h_{i,m},h_{j,n}]=\frac{[A_{i,j}m]_q[cm]_q}{m}q^{-c|m|}
\delta_{m+n,0},
\nonumber\\
&&~[h_i,x_j^\pm(z)]=\pm A_{i,j}x_j^\pm(z),
\nonumber
\\
&&~[h_{i,m}, x_j^+(z)]=\frac{[A_{i,j}m]_q}{m}
q^{-c|m|} z^m x_j^+(z),
\nonumber\\
&&~[h_{i,m}, x_j^-(z)]=-\frac{[A_{i,j}m]_q}{m}
z^m x_j^-(z),
\nonumber
\\
&&(z_1-q^{\pm A_{i,j}}z_2)
x_i^\pm(z_1)x_j^\pm(z_2)
=
(q^{\pm A_{j,i}}z_1-z_2)
x_j^\pm(z_2)x_i^\pm(z_1)~{\rm for}~|A_{i,j}|\neq 0,
\nonumber
\\
&&
~[x_i^\pm(z_1),x_j^\pm(z_2)]=0~{\rm for}~|A_{i,j}|=0,
\nonumber\\
&&~[x_i^+(z_1),x_j^-(z_2)]
=\frac{\delta_{i,j}}{(q-q^{-1})z_1z_2}
\left(
\delta(q^{-c}z_1/z_2)\psi_i^+(q^{\frac{c}{2}}z_2)-
\delta(q^{c}z_1/z_2)\psi_i^-(q^{-\frac{c}{2}}z_2)
\right), \nonumber\\
&& 
\left(
x_i^\pm(z_{1})
x_i^\pm(z_{2})
x_j^\pm(z)-(q+q^{-1})
x_i^\pm(z_{1})
x_j^\pm(z)
x_i^\pm(z_{2})
+x_j^\pm(z)
x_i^\pm(z_{1})
x_i^\pm(z_{2})\right)\nonumber\\
&&+\left(z_1 \leftrightarrow z_2\right)=0
~~~{\rm for}~|A_{i,j}|=1,~i\neq N,\nonumber
\end{eqnarray}
where we have used
$\delta(z)=\sum_{m \in {\bf Z}}z^m$.
Here we have
used the generating function
\begin{eqnarray}
&&x_j^\pm(z)=
\sum_{m \in {\bf Z}}x_{j,m}^\pm z^{-m-1},\nonumber\\
&&
\psi_i^\pm(q^{\pm \frac{c}{2}}z)=q^{\pm h_i}
e^{
\pm (q-q^{-1})\sum_{m>0}h_{i,\pm m}z^{\mp m}}.\nonumber
\end{eqnarray}
}
The relation between two definitions of $U_q(\widehat{sl}(N|1))$ are given by
\begin{eqnarray}
&&h_0=c-(h_{1}+\cdots+h_{N}),~~~
e_i=x_{i,0}^+,~~~f_i=x_{i,0}^-
~~~{\rm for}~~1\leq i \leq N,\nonumber
\\
&&e_0=(-1)[x_{N,0}^- \cdots,[
x_{3,0}^-,[x_{2,0}^-,x_{1,1}^-]_{q^{-1}}]_{q^{-1}}\cdots ]_{q^{-1}}
q^{-h_{1}-h_{2}-\cdots-h_{N}},
\nonumber\\
&&f_0=q^{h_{1}+h_{2}+\cdots+h_{N}}
[\cdots
[[x_{1,-1}^+,x_{2,0}^+]_q, x_{3,0}^+ ]_q, \cdots x_{N,0}^+]_q.
\nonumber
\end{eqnarray}
For instance we have the coproduct as follows.
\begin{eqnarray}
\Delta(h_{i,m})
&=&
h_{i,m}\otimes q^{\frac{cm}{2}}+q^{\frac{3cm}{2}}\otimes 
h_{i,m}~~(m>0),\nonumber\\
\Delta(h_{i,-m})
&=&h_{i,-m}\otimes q^{-\frac{3cm}{2}}+q^{-\frac{cm}{2}}\otimes h_{i,-m}
~~(m>0).\nonumber
\end{eqnarray}

\subsection{Bosonization}

In this section we construct bosonizations of 
quantum superalgebra $U_q(\hat{sl}(N|1))$
for an arbitrary level $k \in {\bf C}$ \cite{K1}. 
We introduce the bosons
and the zero-mode operators
$a_m^j, Q_a^j$ $(m \in {\bf Z},
1\leq j \leq N)$, 
$b_m^{i,j}, Q_b^{i,j}$
$(m \in {\bf Z}, 1\leq i<j \leq N+1)$,
$c_m^{i,j}, Q_c^{i,j}$
$(m \in {\bf Z}, 1\leq i<j \leq N)$
which satisfy
\begin{eqnarray}
&&~[a_m^i,a_n^j]=\frac{[(k+N-1)m]_q[A_{i,j}m]_q}{m}
\delta_{m+n,0},~[a_0^i, Q_a^j]=(k+N-1)A_{i,j},\nonumber
\\
&&~[b_m^{i,j},b_n^{i',j'}]=
-\nu_i \nu_j \frac{[m]_q^2}{m}
\delta_{i,i'}\delta_{j,j'}\delta_{m+n,0},
~[b_0^{i,j},Q_b^{i',j'}]=
-\nu_i \nu_j \delta_{i,i'}\delta_{j,j'},\nonumber
\\
&&~[c_m^{i,j},c_n^{i',j'}]=
\frac{[m]_q^2}{m}
\delta_{i,i'}\delta_{j,j'}
\delta_{m+n,0},
~[c_0^{i,j},Q_c^{i',j'}]=
\delta_{i,i'}\delta_{j,j'},\nonumber
\\
&&~[Q_b^{i,j},Q_b^{i',j'}]=\delta_{j,N+1}\delta_{j',N+1}
\pi \sqrt{-1}~~~~(i,j) \neq (i',j').\nonumber
\end{eqnarray}
Other commutation relations are zero.
In what follows we use the standard symbol of the normal orderings $: :$.
It is convenient to introduce the generating function
$b^{i,j}(z), c^{i,j}(z), b_\pm^{i,j}(z),
a^j_\pm(z)$ and $\left(\frac{\gamma_1}{\beta_1}
\cdots \frac{\gamma_r}{\beta_r}
~a^i \right)\left(z|\alpha \right)$ given by
\begin{eqnarray}
b^{i,j}(z)&=&
-\sum_{m \neq 0}\frac{b_m^{i,j}}{[m]_q}z^{-m}+Q_b^{i,j}
+b_0^{i,j}{\rm log}z,\nonumber
\\
c^{i,j}(z)&=&
-\sum_{m \neq 0}\frac{c_m^{i,j}}{[m]_q}z^{-m}+Q_c^{i,j}+c_0^{i,j}{\rm log}z,\nonumber
\\
b_\pm^{i,j}(z)&=&
\pm (q-q^{-1})\sum_{\pm m>0}b_m^{i,j} 
z^{-m} \pm b_0^{i,j}{\rm log}q,\nonumber\\
a_\pm^{j}(z)&=&
\pm (q-q^{-1})\sum_{\pm m>0}a_m^{j} 
z^{-m}\pm a_0^j {\rm log}q,\nonumber
\\
\left(\frac{\gamma_1}{\beta_1}
\cdots \frac{\gamma_r}{\beta_r}
~a^i \right)\left(z|\alpha \right)&=&
-\sum_{m \neq 0}\frac{[\gamma_1 m]_q \cdots [\gamma_r m]_q}
{[\beta_1 m]_q \cdots [\beta_r m]_q}
\frac{a^i_m}{[m]_q} q^{-\alpha |m|}z^{-m}\nonumber\\
&+&\frac{\gamma_1 \cdots \gamma_r}{\beta_1 \cdots \beta_r}
(Q_a^i+a_0^i {\rm log}z).\nonumber
\end{eqnarray}
In order to avoid divergence we work on the Fock space
defined below.
We introduce the vacuum state $|0\rangle \neq 0$ of 
the boson Fock space by
\begin{eqnarray}
a_m^i|0\rangle=b_m^{i,j}|0\rangle
=c_m^{i,j}|0 \rangle=0~~(m \geq 0).\nonumber
\end{eqnarray}
For $p_a^i \in {\bf C}$ $(1\leq i \leq N)$,
$p_b^{i,j} \in {\bf C}$ $(1\leq i<j \leq N+1)$,
$p_c^{i,j} \in {\bf C}$ $(1\leq i<j \leq N)$,
we set
\begin{eqnarray}
|p_a, p_b, p_c \rangle
&=&e^{
\sum_{i,j=1}^N
\frac{{\rm Min}(i,j)(N-1-{\rm Max}(i,j))}{(N-1)(k+N-1)}
p_a^i Q_a^j}\nonumber\\
&\times&
e^{
-\sum_{1\leq i<j \leq N+1}p_b^{i,j}Q_b^{i,j}
+\sum_{1\leq i<j \leq N}p_c^{i,j}Q_c^{i,j}}|0\rangle.\nonumber
\end{eqnarray}
It satisfies
\begin{eqnarray}
&&
a_0^i|p_a,p_b,p_c\rangle=p_a^i |p_a,p_b,p_c\rangle,~\nonumber\\
&&
b_0^{i,j}|p_a,p_b,p_c\rangle=p_b^{i,j} |p_a,p_b,p_c\rangle,~
c_0^{i,j}|p_a,p_b,p_c\rangle=p_c^{i,j} |p_a,p_b,p_c\rangle.
\nonumber
\end{eqnarray}
The boson Fock space $F(p_a,p_b,p_c)$
is generated by
the bosons $a_m^i, b_m^{i,j}, c_m^{i,j}$
on the vector $|p_a,p_b,p_c\rangle$.
We set the space $F(p_a)$ by
\begin{eqnarray}
F(p_a)=
\bigoplus
_{
p_b^{i,j}=-p_c^{i,j} \in {\bf Z}~(1\leq i<j \leq N)
\atop{
p_b^{i,N+1} \in {\bf Z}~(1\leq i \leq N)
}}F(p_a,p_b,p_c).\nonumber
\end{eqnarray}
We impose the restriction
$p_b^{i,j}=-p_c^{i,j} \in {\bf Z}$ $(1\leq i<j \leq N)$.
We construct 
a bosonization on the space $F(p_a)$.

~\\
{\bf Theorem 3.4}{\it~\cite{K1}~
A bosonization of the quantum superalgebra $U_q(\widehat{sl}(N|1))$
for an arbitrary level $k \in {\bf C}$ is given as follows. 
\begin{eqnarray}
c&=&k \in {\bf C},\nonumber
\\
h_{i}&=&
a_0^i+\sum_{l=1}^i (b_0^{l,i+1}-b_0^{l,i})
+\sum_{l=i+1}^N(b_0^{i,l}-b_0^{i+1,l})
+b_0^{i,N+1}-b_0^{i+1,N+1},
\nonumber\\
h_{N}&=&
a_0^N-\sum_{l=1}^{N-1}(b_0^{l,N}+b_0^{l,N+1}),\nonumber\\
h_{i,m}&=&
q^{-\frac{N-1}{2}|m|}a_m^i+\sum_{l=1}^i
(q^{-(\frac{k}{2}+l-1)|m|}b_m^{l,i+1}-
q^{-(\frac{k}{2}+l)|m|}b_m^{l,i})\nonumber\\
&+&\sum_{l=i+1}^N(q^{-(\frac{k}{2}+l)|m|}b_m^{i,l}-
q^{-(\frac{k}{2}+l-1)|m|}b_m^{i+1,l})\nonumber\\
&+&q^{-(\frac{k}{2}+N)|m|}b_m^{i,N+1}-
q^{-(\frac{k}{2}+N-1)|m|}b_m^{i+1,N+1},\nonumber
\\
h_{N,m}&=&q^{-\frac{N-1}{2}|m|}a_m^N-
\sum_{l=1}^{N-1}(q^{-(\frac{k}{2}+l)|m|}b_m^{l,N}+
q^{-(\frac{k}{2}+l)|m|}b_m^{l,N+1}),
\nonumber\\
x_i^+(z)&=&
\frac{1}{(q-q^{-1})z}
:\sum_{j=1}^i
e^{(b+c)^{j,i}(q^{j-1}z)+\sum_{l=1}^{j-1}
(b_+^{l,i+1}(q^{l-1}z)-b_+^{l,i}(q^lz))}
\times \nonumber
\\
&\times&
\left\{
e^{b_+^{j,i+1}(q^{j-1}z)-
(b+c)^{j,i+1}(q^jz)}-
e^{b_-^{j,i+1}(q^{j-1}z)-
(b+c)^{j,i+1}(q^{j-2}z)}\right\}:,\nonumber
\\
x_N^+(z)
&=&:
\sum_{j=1}^N 
e^{(b+c)^{j,N}(q^{j-1}z)
+b^{j,N+1}(q^{j-1}z)
-\sum_{l=1}^{j-1}(b_+^{l,N+1}(q^lz)+b_+^{l,N}(q^lz))}:,
\nonumber
\\
x_i^-(z)&=&
q^{k+N-1}
:e^{a_+^i(q^{\frac{k+N-1}{2}}z)
-b^{i,N+1}(q^{k+N-1}z)-b_+^{i+1,N+1}(q^{k+N-1}z)+b^{i+1,N+1}(q^{k+N}z)}:
\nonumber
\\
&+&
\frac{1}{(q-q^{-1})z}:
\left\{
\sum_{j=1}^{i-1}
e^{
a_-^i(q^{-\frac{k+N-1}{2}}z)
+(b+c)^{j,i+1}(q^{-k-j}z)
+b_-^{i,n+1}(q^{-k-n}z)-b_-^{i+1,n+1}(q^{-k-n+1}z)}
\right.
\nonumber\\
&\times&
e^{\sum_{l=j+1}^i 
(b_-^{l,i+1}(q^{-k-l+1}z)-b_-^{l,i}(q^{-k-l}z))
+\sum_{l=i+1}^N
(b_-^{i,l}(q^{-k-l}z)-b_-^{i+1,l}(q^{-k-l+1}z))}\nonumber\\
&\times&
\left(
e^{-b_-^{j,i}(q^{-k-j}z)
-(b+c)^{j,i}(q^{-k-j+1}z)}
-
e^{-b_+^{j,i}(q^{-k-j}z)
-(b+c)^{j,i}(q^{-k-j-1}z)}
\right)
\nonumber\\
&+&
e^{a_-^i(q^{-\frac{k+N-1}{2}}z)+(b+c)^{i,i+1}(q^{-k-i}z)}\nonumber\\
&\times& 
e^{
\sum_{l=i+1}^N(b_-^{i,l}(q^{-k-l}z)
-b_-^{i+1,l}(q^{-k-l+1}z))
+b_-^{i,N+1}(q^{-k-N}z)-b_-^{i+1,N+1}(q^{-k-N+1}z)}
\nonumber
\\
&-&
e^{a_+^i(q^{\frac{k+N-1}{2}}z)
+(b+c)^{i,i+1}(q^{k+i}z)}\nonumber\\
&\times&
e^{
\sum_{l=i+1}^N(b_+^{i,l}(q^{k+l}z)
-b_+^{i+1,l}(q^{k+l-1}z))
+b_+^{i,N+1}(q^{k+N}z)-b_+^{i+1,N+1}(q^{k+N-1}z)}
\nonumber\\
&-&
\sum_{j=i+1}^{N-1}
e^{a_+^i(q^{\frac{k+N-1}{2}}z)+
(b+c)^{i,j+1}(q^{k+j}z)}\nonumber\\
&\times&e^{
b_+^{i,N+1}(q^{k+N}z)-b_+^{i+1,N+1}(q^{k+N-1}z)
+\sum_{l=j+1}^N
(b_+^{i,l}(q^{k+l}z)-b_+^{i+1,l}(q^{k+l-1}z))}\nonumber\\
&\times&
\left.
\left(
e^{b_+^{i+1,j+1}(q^{k+j}z)-(b+c)^{i+1,j+1}(q^{k+j+1}z)}
-
e^{b_-^{i+1,j+1}(q^{k+j}z)-(b+c)^{i+1,j+1}(q^{k+j-1}z)}
\right)\right\}:.
\nonumber\\
x_N^-(z)&=&
\frac{1}{(q-q^{-1})z}
:\left\{
\sum_{j=1}^{N-1}
e^{
a_-^N(q^{-\frac{k+N-1}{2}}z)
-b_+^{j,N+1}(q^{-k-j}z)-b^{j,N+1}(q^{-k-j-1}z)}
\right.\nonumber\\
&\times&
e^{
-\sum_{l=j+1}^{N-1}(b_-^{l,N}(q^{-k-l}z)
+b_-^{l,N+1}(q^{-k-l}z))}
\nonumber\\
&\times&
\left.
q^{j-1}\left(
e^{-b_+^{j,N}(q^{-k-j}z)-(b+c)^{j,N}(q^{-k-j-1}z)}
-
e^{
-b_-^{j,N}(q^{-k-j}z)-(b+c)^{j,N}(q^{-k-j+1}z)}
\right)\right\}:
\nonumber\\
&+&
q^{N-1}:\left(
e^{
a_+^N(q^{\frac{k+N-1}{2}}z)-b^{N,N+1}(q^{k+N-1}z)}
-
e^{
a_-^N(q^{-\frac{k+N-1}{2}}z)-b^{N,N+1}(q^{-k-N+1}z)}
\right):.
\nonumber
\end{eqnarray}
}

\subsection{Replacement from $U_q(sl(N|1))$ to $U_q(\widehat{sl}(N|1))$}

In this section we study
the relation between $U_q(sl(N|1))$ and $U_q(\widehat{sl}(N|1))$.
Let us recall the Heisenberg realization
of quantum superalgebra $U_q(sl(N|1))$
\cite{AOS2}.
We introduce the coordinates 
$x_{i,j}$, $(1\leq i<j \leq N+1)$ by
\begin{eqnarray}
x_{i,j}=\left\{\begin{array}{cc}
z_{i,j}&~~~(1\leq i<j \leq N),\\
\theta_{i,j}&~~~(1\leq i \leq N, j=N+1).
\end{array}
\right.
\end{eqnarray}
Here $z_{i,j}$ are complex variables and
$\theta_{i,N+1}$ are the Grassmann odd variables
that satisfy
$\theta_{i,N+1}\theta_{i,N+1}=0$ and
$\theta_{i,N+1}\theta_{j,N+1}=-\theta_{j,N+1}\theta_{i,N+1}$,
$(i \neq j)$.
We introduce the differential operators
$\vartheta_{i,j}=x_{i,j}
\frac{\partial}{\partial x_{i,j}}$, 
$(1\leq i<j \leq N+1)$.

~\\
{\bf Theorem 3.5}{\it~\cite{AOS2}~
We fix parameters $\lambda_i \in {\bf C}$
$(1\leq i \leq N)$.
The Heisenberg realization of $U_q(sl(N|1))$ is given as follows.
\begin{eqnarray}
h_i&=&\sum_{j=1}^{i-1}
(\nu_i \vartheta_{j,i}-\nu_{i+1}\vartheta_{j,i+1})+
\lambda_i-(\nu_i+\nu_{i+1})\vartheta_{i,i+1}
+\sum_{j=i+1}^N
(\nu_{i+1}\vartheta_{i+1,j+1}-\nu_i \vartheta_{i,j+1}),\nonumber\\
e_i&=&\sum_{j=1}^i
\frac{x_{j,i}}{x_{j,i+1}}[\vartheta_{j,i+1}]_q
~q^{
\sum_{l=1}^{j-1}(\nu_i \vartheta_{l,i}-\nu_{i+1}
\vartheta_{l,i+1})},\nonumber\\
f_i&=&\sum_{j=1}^{i-1}
\nu_i \frac{x_{j,i+1}}{x_{j,i}}[
\vartheta_{j,i}]_q q^{\sum_{l=j+1}^{i-1}
(\nu_{i+1}\vartheta_{l,i+1}-\nu_i \vartheta_{l,i})
-\lambda_i+(\nu_i+\nu_{i+1})\vartheta_{i,i+1}
+\sum_{l=i+2}^{N+1}(\nu_i \vartheta_{i,l}-\nu_{i+1}
\vartheta_{i+1,l})}\nonumber\\
&+&x_{i,i+1}\left[\lambda_i-\nu_i \vartheta_{i,i+1}-
\sum_{l=i+2}^{N+1}
(\nu_i \vartheta_{i,l}-\nu_{i+1} \vartheta_{i+1,l})
\right]_q\nonumber\\
&-&
\sum_{j=i+1}^N
\nu_{i+1}\frac{x_{i,j+1}}{x_{i+1,j+1}}
[\vartheta_{i+1,j+1}]_q q^{
\lambda_i+\sum_{l=j+1}^{N+1}
(\nu_{i+1}\vartheta_{i+1,l}-\nu_i \vartheta_{i,l})}.\nonumber
\end{eqnarray}
Here we read $x_{i,i}=1$ and, for Grassmann odd variables
$x_{i,j}$, the expression $\frac{1}{x_{i,j}}$
stands for the derivative $\frac{1}{x_{i,j}}=
\frac{\partial}{\partial x_{i,j}}$.}

We study how to recover the bosonization
of the affine superalgebra $U_q(\widehat{sl}(N|1))$
from the Heisenberg realization of $U_q(sl(N|1))$.
We make the following replacement with suitable argument.
\begin{eqnarray}
\vartheta_{i,j}&\to&-b_\pm^{i,j}(z)/{\rm log}q~~~~~
(1\leq i<j \leq N+1),
\nonumber\\
~[\vartheta_{i,j}]_q&\to&
\left\{
\begin{array}{cc}
\frac{\displaystyle
e^{\pm b_+^{i,j}(z)}-
e^{\pm b_-^{i,j}(z)}}{
\displaystyle
(q-q^{-1})z}
&~~~(j\neq N+1),\\
1&~~~(j=N+1).
\end{array}
\right.\nonumber
\\
x_{i,j}
&\to&
\left\{
\begin{array}{cc}
:e^{(b+c)^{i,j}(z)}:
&~~(j\neq N+1),\\
:e^{-b^{i,j}(z)}:
~{\rm or}~
:e^{-b_\pm^{i,j}(q^{\pm 1}z)-b^{i,j}(z)}:
&~~(j=N+1).
\end{array}
\right.
\nonumber
\\
\lambda_i
&\to&
a_\pm^i(z)/{\rm log}q~~~(1\leq i \leq N),\nonumber
\\
~[\lambda_i]_q
&\to&
\frac{e^{\pm a_+^i(z)}-
e^{\pm a_-^i(z)}}{
\displaystyle (q-q^{-1})z}~~~(1\leq i \leq N).\nonumber
\end{eqnarray}
From the above replacement,
the element $h_i$ of the Heisenberg realization
is replaced as following.
\begin{eqnarray}
q^{h_i}
\to
\left\{
\begin{array}{cc}
e^{a_\pm^i(z)
+\sum_{l=1}^i(b_\pm^{l,i+1}(z)-b_\pm^{l,i}(z))
+\sum_{l=i+1}^N (b_\pm^{i,l}(z)-b_\pm^{i+1,l}(z))}&
~(1\leq i \leq N-1),\\
e^{a_\pm^N(z)-\sum_{l=1}^{N-1}
(b_\pm^{l,N}(z)+b_\pm^{l,N+1}(z))}&~(i=N).
\end{array}
\right.\nonumber
\end{eqnarray}
We impose $q$-shift to variable $z$ of 
the operators $a^i_\pm(z)$,
$b_\pm^{i,j}(z)$. 
For instance, we have to replace 
$a^i_\pm(z) \to a^i_\pm(q^{\pm \frac{c+N-1}{2}}z)$.
Bridging the gap by the $q$-shift,
we have the bosonizations
$\psi_i^\pm(q^{\pm\frac{c}{2}}z) \in U_q(\widehat{sl}(N|1))$ from 
$q^{h_i} \in U_q(sl(N|1))$.
\begin{eqnarray}
\psi_i^\pm(q^{\pm \frac{c}{2}}z)&=&
e^{a_\pm^i(q^{\pm \frac{c+N-1}{2}}z)+
\sum_{l=1}^i(b_\pm^{l,i+1}(q^{\pm(l+c-1)}z)-b_\pm^{l,i}
(q^{\pm(l+c)}z))}\nonumber
\\
&\times&
e^{\sum_{l=i+1}^{N}(b_\pm^{i,l}(q^{\pm(c+l)}z)-
b_\pm^{i-1,l}(q^{\pm(c+l-1)}z))
+b_\pm^{i,N+1}(q^{\pm(c+N)}z)-
b_\pm^{i+1,N+1}(q^{\pm(c+N-1)}z)},\nonumber
\\
\psi_N^\pm(q^{\pm \frac{c}{2}}z)&=&
e^{a_\pm^N(q^{\pm \frac{c+N-1}{2}}z)-
\sum_{l=1}^{N-1}
(b_\pm^{l,N}(q^{\pm (c+l)}z)
+b_\pm^{l,N+1}(q^{\pm (c+l)}z))}.\nonumber
\end{eqnarray}
In this replacement, one element $q^{h_i}$
goes to two elements $\psi_i^\pm(q^{\pm \frac{c}{2}}z)$.
Hence this replacement is not a map.
Replacements from $e_i, f_i$ to $x_i^\pm(z)$ are given by similar way,
however they are more complicated. See details in \cite{K1}.

\subsection{Wakimoto Realization}

In this section we give the Wakimoto realization
${\cal F}(p_a)$ whose character coincides with those of 
the Verma module \cite{K2}.
We introduce the operators $\xi_m^{i,j}$ and
$\eta_m^{i,j}$ $(1\leq i<j \leq N, m \in {\bf Z})$ by
\begin{eqnarray}
\eta^{i,j}(z)=\sum_{m \in {\bf Z}}\eta_{m}^{i,j}
z^{-m-1}=:e^{c^{i,j}(z)}:,~~
\xi^{i,j}(z)=\sum_{m \in {\bf Z}}\xi_{m}^{i,j}z^{-m}=:e^{-c^{i,j}(z)}:.
\nonumber
\end{eqnarray}
The Fourier components $\eta_m^{i,j}
=\oint \frac{dz}{2\pi \sqrt{-1}}z^m \eta^{i,j}(z)$,
$\xi_m^{i,j}=
\oint \frac{dz}{2\pi \sqrt{-1}}z^{m-1}\xi^{i,j}(z)$ $(m \in {\bf Z})$
are well defined on the space ${F}(p_a)$.
We focus our attention on the operators 
$\eta_0^{i,j}, \xi_0^{i,j}$ satisfying
$(\eta_0^{i,j})^2=0$, $(\xi_0^{i,j})^2=0$.
They satisfy
\begin{eqnarray}
{\rm Im} (\eta_0^{i,j})={\rm Ker} (\eta_0^{i,j}),~
{\rm Im} (\xi_0^{i,j})={\rm Ker} (\xi_0^{i,j}),
~\eta_0^{i,j}\xi_0^{i,j}+\xi_0^{i,j}\eta_0^{i,j}=1.
\nonumber
\end{eqnarray}
We have a direct sum decomposition.
\begin{eqnarray}
&&F(p_a)=
\eta_0^{i,j}\xi_0^{i,j}F(p_a) \oplus 
\xi_0^{i,j}\eta_0^{i,j}F(p_a),
\nonumber\\
&&{\rm Ker} (\eta_0^{i,j})=
\eta_0^{i,j}\xi_0^{i,j}F(p_a),~
{\rm Coker} (\eta_0^{i,j})=
\xi_0^{i,j}\eta_0^{i,j}F(p_a)
=F(p_a)/ 
(\eta_0^{i,j}\xi_0^{i,j})F(p_a).
\nonumber
\end{eqnarray}
We set the operator $\eta_0, \xi_0$ by
\begin{eqnarray}
\eta_0=\prod_{1\leq i < j \leq N}\eta_0^{i,j},~~~
\xi_0=\prod_{1\leq i < j \leq N}\xi_0^{i,j}.
\nonumber
\end{eqnarray}
~\\
{\bf Definition 3.6}{\it~\cite{K2}~
We introduce the subspace ${\cal F}(p_a)$ by 
\begin{eqnarray}
{\cal F}(p_a)
=\eta_0 \xi_0
F(p_a).\nonumber
\end{eqnarray}
We call ${\cal F}(p_a)$ the Wakimoto realization.}

\section{Screening and Vertex Operator}

In this section we give the screening that
commutes with the quantum superalgebra
$U_q(\widehat{sl}(N|1))$.
We propose the vertex operators and the correlation functions.

\subsection{Screening}

In this section we give the screening ${\cal Q}_i$ $(1\leq i \leq N)$
that commutes with the quantum superalgebra
$U_q(\widehat{sl}(N|1))$
for an arbitrary level $k \neq -N+1$ \cite{K3}.
The Jackson integral with parameter $p \in {\bf C}$ $(|p|<1)$
and $s \in {\bf C}^*$ is defined by
\begin{eqnarray}
\int_0^{s \infty}f(z)d_pz=s(1-p)\sum_{m \in {\bf Z}}f(sp^m)p^m.
\nonumber
\end{eqnarray}
In order to avoid divergence we work in the Fock space.

~\\
{\bf Theorem 4.1}{\it~\cite{K3}~
The screening ${\cal Q}_i$ commutes with the quantum superalgebra.
\begin{eqnarray}
[{\cal Q}_i, U_q(\widehat{sl}(N|1))]=0~~~(1\leq i \leq N).\nonumber
\end{eqnarray}
We have introduced the screening operators 
${\cal Q}_i$ $(1\leq i \leq N)$ as follows.
\begin{eqnarray}
{\cal Q}_i=\int_0^{s \infty}
:e^{-\left(\frac{1}{k+N-1}a^i\right)
\left(z\left|\frac{k+N-1}{2}\right.\right)}
\widetilde{S}_i(z):d_pz,~~~(p=q^{2(k+N-1)}).\nonumber
\end{eqnarray}
Here we have set
the bosonic operators 
$\widetilde{S}_i(z)$ $(1\leq i \leq N)$ by
\begin{eqnarray}
\widetilde{S}_i(z)&=&\frac{1}{(q-q^{-1})z}\sum_{j=i+1}^N
:\left(
e^{-b_-^{i,j}(q^{N-1-j}z)-(b+c)^{i,j}(q^{N-j}z)}
-
e^{-b_+^{i,j}(q^{N-1-j}z)-(b+c)^{i,j}(q^{N-j-2}z)}\right)\nonumber\\
&^\times&
e^{(b+c)^{i+1,j}(q^{N-1-j}z)+\sum_{l=j+1}^N
(b_-^{i+1,l}(q^{N-l}z)-b_-^{i,l}(q^{N-l-1}z))
+b_-^{i+1,N+1}(z)-b_-^{i,N+1}(q^{-1}z)}:
\nonumber\\
&+&q:e^{b^{i,N+1}(z)
+b_+^{i+1,N+1}(z)-b^{i+1,N+1}(qz)}:
~~~~~(1\leq i \leq N-1),
\nonumber\\
\widetilde{S}_N(z)&=&-q^{-1}:e^{b^{N,N+1}(z)}:.\nonumber
\end{eqnarray}
}

\subsection{Vertex Operator}

In this section we introduce the vertex operators
$\Phi(z)$, $\Phi^*(z)$ \cite{K3}.
Let ${\cal F}$ and ${\cal F}'$ be
$U_q(\widehat{sl}(N|1))$ representation for an arbitrary level $k \neq -N+1$.
Let $V_{\alpha}$ and $V_{\alpha}^{* S}$ be
$2^N$-dimensional typical representation with a parameters $\alpha$
\cite{PT}.
Let $\{v_j\}_{j=1}^{2^N}$ be the basis of $V_{\alpha}$.
Let $\{v_j^*\}_{j=1}^{2^N}$ be the dual basis of $V_{\alpha}^{* S}$,
satisfying $(v_i|v_j^*)=\delta_{i,j}$.
Let $V_{\alpha, z}$ and $V_{\alpha,z}^{* S}$ be
the evaluation module and its dual
of the typical representation.
For instance,
the $8$-dimensional representation 
$V_{\alpha, z}$ of $U_q(\widehat{sl}(3|1))$
is given by
\begin{eqnarray}
h_1&=&E_{3,3}-E_{4,4}+E_{5,5}-E_{6,6},\nonumber\\
h_2&=&E_{2,2}-E_{3,3}+E_{6,6}-E_{7,7},\nonumber\\
h_3&=&\alpha(E_{1,1}+E_{2,2})+(\alpha+1)(E_{3,3}+E_{4,4}+E_{5,5}+E_{6,6})+
(\alpha+2)(E_{7,7}+E_{8,8}),\nonumber\\
e_1&=&E_{3,4}+E_{5,6},\nonumber\\
e_2&=&E_{2,3}+E_{6,7},\nonumber\\
e_3&=&\sqrt{[\alpha]_q}E_{1,2}-\sqrt{[\alpha+1]_q}(E_{3,5}+E_{4,6})
+\sqrt{[\alpha+2]_q}E_{7,8},\nonumber\\
f_1&=&E_{4,3}+E_{6,5},\nonumber\\
f_2&=&E_{3,2}+E_{7,6},\nonumber\\
f_3&=&\sqrt{[\alpha]_q}E_{2,1}-\sqrt{[\alpha+1]_q}(E_{5,3}+E_{6,4})+
\sqrt{[\alpha+2]_q}E_{8,7},\nonumber\\
h_0&=&-\alpha(E_{1,1}+E_{4,4})
-(\alpha+1)(E_{2,2}+E_{3,3}+E_{6,6}+E_{7,7})
-(\alpha+2)(E_{5,5}+E_{8,8}),\nonumber\\
e_0&=&
-z(\sqrt{[\alpha]_q}E_{4,1}
-\sqrt{[\alpha+1]_q}(E_{6,2}+E_{7,3})
+\sqrt{[\alpha+2]_q}E_{8,5}),\nonumber\\
f_0&=&
z^{-1}(\sqrt{[\alpha]_q}E_{1,4}
-\sqrt{[\alpha+1]_q}(E_{2,6}+E_{3,7})
+\sqrt{[\alpha+2]_q}E_{5,8}).\nonumber
\end{eqnarray}
Consider the following intertwiners 
of $U_q(\widehat{sl}(N|1))$-representation \cite{FR}.
\begin{eqnarray}
\Phi(z): {\cal F} \longrightarrow {\cal F}' \otimes V_{\alpha,z},~~~
\Phi^*(z): {\cal F} \longrightarrow {\cal F}' \otimes V_{\alpha,z}^{* S}.
\nonumber
\end{eqnarray}
They are intertwiners in the sense that for any 
$x \in U_q(\widehat{sl}(N|1))$,
\begin{eqnarray}
\Phi(z)\cdot x=\Delta(x) \cdot \Phi(z),~~~
\Phi^*(z)\cdot x=\Delta(x) \cdot \Phi^*(z).\nonumber
\end{eqnarray}
We expand the intertwining operators.
\begin{eqnarray}
\Phi(z)=\sum_{j=1}^{2^N}\Phi_j(z)\otimes v_j,~~~
\Phi^*(z)=\sum_{j=1}^{2^N}\Phi_j^*(z)\otimes v_j^*.\nonumber
\end{eqnarray}
We set the ${\bf Z}_2$-grading of 
the intertwiner be $|\Phi(z)|=|\Phi^*(z)|=0$.
For $l_a=(l_a^1,l_a^2,\cdots,l_a^N) \in {\bf C}^N$ and 
$\beta \in {\bf C}$,
we set the bosonic operator $\phi^{ l_a}(z|\beta)$ by
\begin{eqnarray}
\phi^{ l_a}(z|\beta)=:e^{
\sum_{i,j=1}^N \left(\frac{l_a^i}{k+N-1}
\frac{{\rm Min}(i,j)}{N-1}\frac{N-1-{\rm Max}(i,j)}{1}
a^j\right)(z|\beta)}:.\nonumber
\end{eqnarray}
In order to balance thegbackground chargeh 
of the vertex operators, we introduce the product of the screenings
${\cal Q}^{(t)}$ for 
$t=(t_1,t_2,\cdots,t_N) \in {\bf N}^N$. 
\begin{eqnarray}
{\cal Q}^{(t)}={\cal Q}_1^{t_1} {\cal Q}_2^{t_2} 
\cdots {\cal Q}_N^{t_N}.\nonumber
\end{eqnarray}
The screening operator ${\cal Q}^{(t)}$ give rise to the map,
\begin{eqnarray}
{\cal Q}^{(t)} : {\cal F}(p_a)\to {\cal F}(p_a+\hat{t}).\nonumber
\end{eqnarray}
Here 
$\hat{t}=(\hat{t}_1,\hat{t}_2,\cdots,\hat{t}_N)$ where $\hat{t}_i=
\sum_{j=1}^N A_{i,j}t_j$.

~\\
{\bf Theorem 4.2}{\it~\cite{K3}~
For $k=\alpha\neq 0,-1,-2,\cdots,-N+1$,
bosonizations
of the special components of the vertex operators
$\Phi^{(t)}(z)$ and $\Phi^{* (t)}(z)$
are given by
\begin{eqnarray}
\Phi_{2^N}^{(t)}(z)&=&{\cal Q}^{(t)}
\phi^{\hat{l}}\left(q^{k+N-1} z\left|-\frac{k+N-1}{2}\right.
\right),\nonumber\\
\Phi_1^{* (t)}(z)&=&{\cal Q}^{(t)}
\phi^{\hat{l}^*}\left(q^{k}z
\left|-\frac{k+N-1}{2}\right.\right),\nonumber
\end{eqnarray}
where 
we have used
$\hat{l}=-(0,\cdots,0,\alpha+N-1)$,
$\hat{l}^*=(0,\cdots,0,\alpha)$ and $t=(t_1,t_2,\cdots,t_N) \in {\bf N}^N$.
The other components $\Phi_j^{(t)}(z)$ and $\Phi_j^{*(t)}(z)$ 
$(1\leq j \leq 2^N)$
are determined by the intertwining property
and are represented 
by multiple contour integrals of Drinfeld currents
and the special components $\Phi_{2^N}^{(t)}(z)$ and $\Phi_1^{*(t)}(z)$.
We have checked this theorem
for $N=2,3,4$.}

~\\
Here we give additional explanation on the above theorem.
The explicit formulae of
the intertwining properties 
$\Phi^{(t)}(z)\cdot x=\Delta(x)\cdot \Phi^{(t)}(z)$ for 
$U_q(\widehat{sl}(3|1))$
are summarized as follows.
We have set the ${\bf Z}_2$-grading of $V_\alpha$ as follows :
$|v_1|=|v_5|=|v_6|=|v_7|=0$, and 
$|v_2|=|v_3|=|v_4|=|v_8|=1$.
\begin{eqnarray}
\Phi_3^{(t)}(z)&=&[\Phi_4^{(t)}(z),f_1]_q,~
\Phi_5^{(t)}(z)=
[\Phi_6^{(t)}(z),f_1]_q,
\nonumber\\
\Phi_2^{(t)}(z)&=&[\Phi_3^{(t)}(z),f_2]_q,~
\Phi_6^{(t)}(z)=[\Phi_7^{(t)}(z),f_2]_q,
\nonumber\\
\Phi_1^{(t)}(z)&=&
\frac{1}{\sqrt{[\alpha]_q}}
[\Phi_2^{(t)}(z),f_3]_{q^{-\alpha}},~
\Phi_3^{(t)}(z)=
\frac{-1}{\sqrt{[\alpha+1]_q}}
[\Phi_5^{(t)}(z),f_3]_{q^{-\alpha-1}},\nonumber
\\
\Phi_4^{(t)}(z)&=&
\frac{-1}{\sqrt{[\alpha+1]_q}}
[\Phi_6^{(t)}(z),f_3]_{q^{-\alpha-1}},~
\Phi_7^{(t)}(z)=
\frac{1}{\sqrt{[\alpha+2]_q}}
[\Phi_8^{(t)}(z),f_3]_{q^{-\alpha-2}}.\nonumber
\end{eqnarray}
The elements $f_j$ are written by
contour integral of the Drinfeld current
$f_j =\oint \frac{dw}{2\pi \sqrt{-1}}x_j^-(w)$.
Hence the components $\Phi_j^{(t)}$ $(1\leq j \leq 8)$
are represented 
by multiple contour integrals of Drinfeld currents
$x_j^-(w)$ $(1\leq j \leq 3)$
and the special component $\Phi_{8}^{(t)}(z)$.

\subsection{Correlation Function}

In this section we study the correlation function 
as an application of
the vertex operators .
We study non-vanishing property of
the correlation function
which is defined to be the trace of the vertex operators
over the Wakimoto module of $U_q(\widehat{sl}(N|1))$.
We propose 
the $q$-Virasoro operator $L_0$ for $k=\alpha \neq -N+1$
as follows.
\begin{eqnarray}
L_0&=&\frac{1}{2}\sum_{i,j=1}^N \sum_{m \in {\bf Z}}
:a_{-m}^i\frac{m^2
[{\rm Min}(i,j)m]_q
[(N-1-{\rm Max}(i,j))m]_q
}{[m]_q[(k+N-1)m]_q[(N-1)m]_q[m]_q}a_m^j:\nonumber\\
&&+\sum_{i,j=1}^N \frac{{\rm Min}(i,j)(N-1-{\rm Max}(i,j))}
{(k+N-1)(N-1)}a_0^j\nonumber\\
&&-\frac{1}{2}\sum_{1\leq i<j \leq N}\sum_{m \in {\bf Z}}
:b_{-m}^{i,j}\frac{m^2}{[m]_q^2}b_m^{i,j}:
+\frac{1}{2}\sum_{1\leq i<j \leq N}
\sum_{m \in {\bf Z}}:c_{-m}^{i,j}\frac{m^2}{[m]_q^2}c_m^{i,j}:
\nonumber\\
&&+\frac{1}{2}\sum_{1\leq i \leq N}\sum_{m \in {\bf Z}}
:b_{-m}^{i,N+1}\frac{m^2}{[m]_q^2}b_m^{i,N+1}:
+\frac{1}{2}\sum_{1\leq i \leq N}b_{0}^{i,N+1}.
\nonumber
\end{eqnarray}
The $L_0$ eigenvalue of $|l_a,0,0\rangle$ is 
$\frac{1}{2(k+N-1)}(\bar{\lambda}|\bar{\lambda}+2\bar{\rho})$,
where $\bar{\rho}=\sum_{i=1}^N \bar{\Lambda}_i$ and $\bar{\lambda}=
\sum_{i=1}^N l_a^i \bar{\Lambda}_i$.

~\\
{\bf Theorem 4.3}{\it~\cite{K3}~
For $k=\alpha \neq 0,-1,-2,\cdots,-N+1$, 
the correlation function of
the vertex operators,
\begin{eqnarray}
{\rm Tr}_{{\cal F}(l_a)}\left(q^{L_0}
{\Phi}_{i_1}^{*(y_{(1)})}(w_1)
\cdots 
{\Phi}_{i_m}^{*(y_{(m)})}(w_m)
{\Phi}_{j_1}^{(x_{(1)})}(z_1)
\cdots 
{\Phi}_{j_n}^{(x_{(n)})}(z_n)
\right)\neq 0,\nonumber
\end{eqnarray}
if and only if
$x_{(s)}=(x_{(s),1},x_{(s),2},\cdots,x_{(s),N}) \in {\bf N}^N$ $(1\leq s \leq n)$ 
and $y_{(s)}=(y_{(s),1},y_{(s),2},\cdots,y_{(s),N})\in {\bf N}^N$ 
$(1\leq s \leq m)$ satisfy the following condition.
\begin{eqnarray}
\sum_{s=1}^n x_{(s),i}+\sum_{s=1}^m y_{(s),i}=
\frac{(n-m) i}{N-1}\alpha+n \cdot i~~~~~(1\leq i \leq N).\nonumber
\end{eqnarray}
}
~\\~\\
{\bf Acknowledgements.}\\
This work is supported by the Grant-in-Aid for
Scientific Research {\bf C} (21540228)
from Japan Society for Promotion of Science. 
The author would like to thank
Professor Branko Dragovich
and the organizing committee of the $7$-th Mathematical Physics Meeting
for an invitation.

\begin{appendix}

\end{appendix}

\end{document}